\documentclass[12pt]{article}
\usepackage{amssymb,amsmath,epsfig}

\begin{document}

\title{\bf Study of Center of Mass Energy by Particles Collision in Some Black Holes}
\author{M. Sharif \thanks{msharif.math@pu.edu.pk} and Nida Haider
\thanks{nida.haider12@gmail.com}\\
Department of Mathematics, University of the Punjab,\\
Quaid-e-Azam Campus, Lahore-54590, Pakistan.}

\date{}

\maketitle
\begin{abstract}
This paper is devoted to the study of particles collision for two
well-known black holes. We consider particles moving in equatorial
plane and calculate their center of mass energy. Firstly, we explore
center of mass energy of a regular black hole. In this case,
acceleration and collision of particles lead to high center of mass
energy which is independent of event horizon and naked singularity.
Secondly, we investigate the center of mass energy of Plebanski and
Demianski black hole (non-extremal) with zero NUT parameter. Here
the center of mass energy depends upon the rotation parameter. We
conclude that the center of mass energy becomes infinitely large for
both black holes.
\end{abstract}

\section{Introduction}

Particle accelerators are devices that propel and accelerate charged
particles (like protons) to high speed. Physicists use them to study
the nature of matter and energy. Tevatron and Large Hadron Collider
are the particle accelerators which propel and produce collision
between particles at the center of mass energy (CME) upto 10TeV. In
particles collision, the CME is the energy required for the creation
of new particles. The fascinating possibility to study this energy
is to make use of naturally occurring processes in the vicinity of
astrophysical objects. Black hole (BH) can behave as a particle
accelerator and can accelerate the colliding particles to an
unlimited CME. The CME of colliding particles can grow limitlessly
in different situations, e.g., nature of colliding particles,
BH/naked singularity, modified gravity theories etc. When two
particles collide near horizon, the energy grows infinitely in the
CM frame because particles are infinitely blueshifted near the
horizon.

Banados, Silk and West (2009) discussed (BSW effect) the effect of
infinite growth of energy in the CM frame due to collision of
particles near the horizon. It was first studied in the locality of
the Kerr (extremal) BH and was shown that two colliding particles
propelling in equatorial plane can have arbitrarily large CME near
the maximal BH spin. Lake (2010) explored the effect of particles
collision near the horizon of the Kerr (non-extremal) BH and
obtained finite CME. Grib and Pavlov (2011) suggested that CME can
become infinite in the non-extremal case if two scattering particles
with equal masses collide close to the horizon. The BSW effect has
also been studied for Sen BH (Wei et al. 2010), Kerr-Taub-NUT BH
(Liu et al. 2011) and Kaluza-Klein BH. Mao et al. (2011) found
infinitely large CME for particles colliding at the horizon of
charged non-rotating and extremal rotating Kaluza-Klein BH.

Piran et al. (1975, 1977) investigated collisions with infinite CME
for two particles colliding in energy extraction process - known as
collisional Penrose. Bejger et al. (2012)  studied this process near
the horizon of (extremal) Kerr BH. Wei et al. (2010) discussed the
effect of charge on the CME for stringy and Kerr-Newman BHs and
obtained arbitrarily large CME. Harada et al. (2012) found this
energy for colliding particles near the horizon of a BH with maximal
rotation and concluded that it is arbitrarily large for the critical
particles (with fine tuned angular momentum). Joshi and Patil
(2011a, 2011b, 2012a) explored Reissner-Nordstrom (RN) and Kerr BHs
and found high CME in the naked singularity. They concluded that the
high energy collision can also be seen in the BH having naked
singularity with no event horizon. The same authors (2012b)
described that the high energy collision can take place near the
naked singularity of Janis Newmann Winicor metric (found by adding a
massless scalar field in the Schwarzschild BH), while these
collisions are not present in the Schwarzschild BH. They also proved
that BH having no event horizon or singularity can also have high
CME for particular values of parameters $m$ and $q$ (2012c). Hussain
(2012) found that the CME of colliding particles in Plebanski and
Demianski BH (extremal) with zero NUT parameter is unlimited at the
acceleration and event horizons.

Zaslavskii (2010a, 2010b, 2011)  studied the universal property for
particle acceleration and generalized the BSW effect for dirty BHs
with non-equatorial motion of colliding particles. Harada and Kimura
(2011) studied the collision in non-equatorial plane with an
unboundedly high CME for (extremal) Kerr BH. Yao et al. (2012)
explored the CME for the collision of particles with non-equatorial
motion in Kerr-Newman BH and discussed the effect of acceleration.
Liu et al. (2011) obtained arbitrarily high CME for colliding
particles with non-equatorial motion near horizon of (extremal)
Kerr-Newman BH. Jacobson and Sotirious (2010) showed that it takes
an infinite time to attain infinite CME.

In this paper, we study CME by particles collision in equatorial
plane for two well-known BHs. The paper is organized as follows. In
sections \textbf{2} and \textbf{3}, we explore CME for regular BH
and for charged accelerating and rotating BH near the horizon. The
last section summarizes the results.

\section{Particle Acceleration in Regular Black Hole}

This section is devoted to study the particle acceleration in a
static spherically symmetric BH coupled to a non-linear
electrodynamics. The general form of a regular BH is
\begin{equation}\label{1}
ds^{2}=-Y(r)dt^{2}+\frac{1}{Y(r)}dr^{2}+r^{2}d\Omega^{2}_{2},
\end{equation}
where $d\Omega^{2}_{2}=d\theta^{2}+\sin^{2}\theta d\phi^{2}$ is a
$2$-dimensional sphere and
\begin{equation}\label{2}
Y(r)=1-\frac{2M(r)}{r}.
\end{equation}
Ayon and Garcia (2000) proposed a non-singular BH solution coupled
with nonlinear electrodynamics whose line element is
\begin{eqnarray}\nonumber
ds^{2}&=&-(1-\frac{2mr^{2}}{(r^{2}+q^{2})^{\frac{3}{2}}}
+\frac{q^{2}r^{2}}{(r^{2}+q^{2})^{2}})dt^{2}+(1-\frac{2mr^{2}}{(r^{2}+q^{2})^{\frac{3}{2}}}
\\\label{3}
&+&\frac{q^{2}r^{2}}{(r^{2}+q^{2})^{2}})^{-1}dr^{2}+r^{2}d\Omega^{2}_{2},
\end{eqnarray}
for which the electric field is
\begin{equation}\label{4}
E(r)= qr^{4}(\frac{r^{2}-5q^{2}}{(r^{2}+q^{2})^{4}}+
\frac{15m}{2(r^{2}+q^{2})^{\frac{7}{2}}}).
\end{equation}
Here $m$ and $q$ are the mass and magnetic charge parameter of BH,
respectively. Comparing Eqs.(\ref{1}) and (\ref{3}), we have
\begin{equation*}
Y(r)=1-\frac{2mr^{2}}{(r^{2}+q^{2})^{\frac{3}{2}}}
+\frac{q^{2}r^{2}}{(r^{2}+q^{2})^{2}},
\end{equation*}
leading to
\begin{equation*}
M(r)=\frac{mr^{3}}{(r^{2}+q^{2})^{\frac{3}{2}}}+\frac{q^{2}r^{3}}{2(r^{2}+q^{2})^{2}}.
\end{equation*}
Expanding by Taylor's expansion at $r=0$, i.e., near the center, it
follows that
\begin{equation*}
M(r)=(\frac{m}{q^{3}}-\frac{1}{2q^{2}})r^{3}-(\frac{3m}{2q^{5}}-\frac{1}{q^{4}})r^{5}+O(r^{7}).
\end{equation*}
The general expression is
\begin{equation*}
M(r)=M_{0}+M_{1}r+M_{2}r^{2}+M_{3}r^{3}+\cdots.
\end{equation*}
Comparing both these equations, we obtain $M_{0}=M_{1}=M_{2}=0$,
while $M_{3}\neq0$. Using the condition (Joshi and Patil 2012c), it
follows that the center is non-singular. Moreover, this solution
asymptotically behaves as RN solution, i.e.,
\begin{equation*}
g_{00}=1+\frac{2m}{r}+\frac{q^{2}}{r^{2}}+O(\frac{1}{r^{3}}).
\end{equation*}
Thus we can write the metric near the center as
\begin{equation}\label{5}
ds^{2}=-(1-2M_{3}r^{2})dt^{2}+(1-2M_{3}r^{2})^{-1}dr^{2}+r^{2}d\Omega_{2}^{2}.
\end{equation}

Consider a particle exhibiting geodesic motion in a static
spherically symmetric spacetime. Let
$U^{a}=(U^{t},U^{r},U^{\theta},U^{\phi})$ be the four velocity of
the particle, which is restricted to equatorial motion
$(\theta=\frac{\pi}{2})$, hence $U^{\theta}=0$. Thus the metric
admits only two Killing vectors $\partial_{\phi}$ and
$\partial_{t}$. Using these Killing vectors, we can  define energy
and angular momentum by
\begin{eqnarray*}
E&=&-g_{ab}(\partial_{t})^{a}U^{b}=-g_{tt}U^{t}-g_{t\phi}U^{\phi},\\
L&=&g_{ab}(\partial_{\phi})^{a}U^{b}=g_{t\phi}U^{t}+g_{\phi\phi}U^{\phi},
\end{eqnarray*}
which are interpreted as constants of motion, i.e., energy and
angular momentum are conserved throughout the motion. Using
Eq.(\ref{1}), these quantities turn out to be
\begin{equation*}
E=Y(r)U^{t}, \quad L=r^{2}U^{\phi},
\end{equation*}
leading to
\begin{equation}\label{6}
U^{t}=\frac{E}{Y(r)}, \quad U^{\phi}=\frac{L}{r^{2}}.
\end{equation}
Using normalization condition, $g_{ab}U^{a}U^{b}=-1$, the radial
component becomes
\begin{equation}\label{7}
U^{r}=\pm[E^{2}-Y(r)(1+\frac{L^{2}}{r^{2}})]^{\frac{1}{2}}.
\end{equation}
Here positive and negative signs correspond to ingoing and outgoing
particles along the radial direction. We define the effective
potential along radial direction as
\begin{equation}\label{8}
V_{eff}(r)= Y(r)(1+\frac{L^{2}}{r^{2}}).
\end{equation}
Thus Eq.(\ref{7}) leads to
\begin{equation}\label{9}
(U^{r})^{2}+V_{eff}(r)=E^{2}.
\end{equation}
The quantity $U^{r}\geq0$ provides the necessary condition for a
particle to reach the center. When velocity is positive, particle
reaches the center but it turns back for zero velocity.
\begin{figure}
\centering \epsfig{file=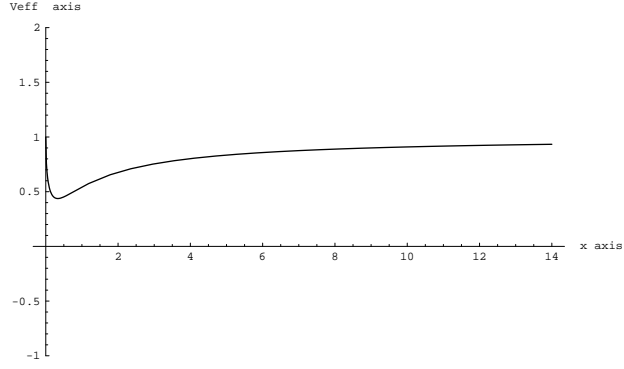,width=.60\linewidth}\caption{The
effective potential is plotted against $x=\frac{r}{m}$.}
\end{figure}

Now we evaluate the minimum of $Y(r)$ graphically as shown in Figure
\textbf{1}. It is observed that the curve admits a minimum at
$x=0.333$, which gives $r_{min}=0.6q$ and the corresponding value of
$Y(r)$ is
\begin{equation}\label{10}
Y(r_{min})= 1.19-0.453\frac{m}{q}.
\end{equation}
When $m>2.6q$, we have $Y(r_{min})<0$. Moreover, we see that
$Y(r=0)=1$ and $Y(r)\rightarrow1$ as $r\rightarrow\infty$. This
implies that the function admits a zero for two values of $r$, i.e.,
$Y(r)=0$ for $r=r_{1}$ (say) and $r=r_{2}$ (say), where
$0<r_{1}<r_{min}$ and $r_{min}<r_{2}<\infty$. In this case, the
metric has its inner and outer horizons at $r=r_{1}$ and $r=r_{2}$,
respectively. For these values of mass and charge, the inner and
outer horizons are defined by $-k_{\mu}k^{\mu}=Y(r)=0$. In this case
metric corresponds to BH.

The curvature invariants $R_{1}=R_{ab}g^{ab}$, $R_{2}=R_{ab}R^{ab}$
and $K=R_{abcd}R^{abcd}$, yield
\begin{eqnarray*}
R_{1}&=&\frac{1}{(q^{2}+r^{2})^{\frac{9}{2}}}
[6q^{2}m(4q^{4}+3r^{2}q^{2}-r^{4})
+12q^{4}(q^{2}-r^{2})(q^{2}+r^{2})^{\frac{1}{2}}],
\\R_{2}&=&\frac{1}{(q^{2}+r^{2})^{8}}[2q^{4}\{18q^{8}+q^{2}r^{4}(81m^{4}
-20r^{2}-168m(q^{2}+r^{2})^{\frac{1}{2}})-36q^{6}\\&\times&(-2m^{2}+r^{2}+2m\sqrt{q^{2}+r^{2}}
)+r^{6}(117m^{2}+2r^{2}+30m\sqrt{q^{2}+r^{2}})\}],
\\K&=&\frac{4}{(q^{2}+r^{2})^{8}}[6q^{12}+12m^{2}r^{10}
-24q^{2}r^{8}(m+\sqrt{q^{2}+r^{2}})+q^{6}r^{4}(129m^{2}\\&-&44r^{2}
-192m\sqrt{q^{2}+r^{2}})-12q^{10}(-2m^{2}+r^{2}+2m\sqrt{q^{2}+r^{2}})
+2q^{8}r^{2}\\&\times&(6m^{2}+34r^{2}+15m\sqrt{q^{2}+r^{2}})
+q^{4}r^{6}(105m^{2}+14r^{2}+90m\sqrt{q^{2}+r^{2}})].
\end{eqnarray*}
Here $R_{1}$, $R_{2}$ and $K$ are finite for finite values of $r$
which show that all of them are bounded everywhere. Thus for
$m>2.6q$, the singularities appearing in the metric are only
coordinate singularities describing the existence of horizon. As all
the curvature invariants are finite everywhere and as
$r\rightarrow\infty,~2M(r)\rightarrow2m$, which gives $2M(r)<r$,
thus for $Y(r_{min})>0$, we neither have horizon nor singularity. In
this case, the conditions $Y(0)=1$ and $Y'(0)=0$ are also satisfied.

When $q=0.4m$, we have $Y(r_{min})=0$ and hence both inner and outer
horizons coincide. In other words, these shrink into a single
horizon, i.e., $r_{min}=r_{1}=r_{2}$, which can be related to
extremal BH as in the RN solution. When charge and mass satisfy the
constraint $q>0.4m$, we obtain $Y(r_{min})>0$, which is the
condition for the avoidance of horizon (Joshi and Patil 2012c). Thus
we can avoid the singularities and event horizons.

Now we investigate CME of two particles having masses $m_{1}$ and
$m_{2}$. In terms of four-momentum
$p^{a}_{i},~(i=1,2,~a=t,r,\theta,\phi)$, the CME of two particles is
(Liu et al. 2011)
\begin{equation*}
E^{2}_{cm}=-p^{a}_{i}p_{ai},
\end{equation*}
which yields
\begin{equation}\label{11}
E^{2}_{cm}=2m_{1}m_{2}[\frac{(m_{1}-m_{2})^{2}}
{2m_{1}m_{2}}+(1-g_{ab}U^{a}_{1}U^{b}_{2})].
\end{equation}
When $m_{1}=m_{2}=m$, using Eq.(\ref{3}), it follows that
\begin{eqnarray}\nonumber
\frac{E^{2}_{cm}}{2m^{2}}&=&1+\frac{E_{1}E_{2}}{(1-\frac{2mr^{2}}
{(r^{2}+q^{2})^{\frac{3}{2}}}+\frac{q^{2}r^{2}}{(r^{2}+q^{2})^{2}})}
-\frac{1}{(1-\frac{2mr^{2}}{(r^{2}+q^{2})^{\frac{3}{2}}}
+\frac{q^{2}r^{2}}{(r^{2}+q^{2})^{2}})}\\\nonumber
&\times&[E_{1}^{2}-(1-\frac{2mr^{2}}{(r^{2}+q^{2})^{\frac{3}{2}}}+
\frac{q^{2}r^{2}}{(r^{2}+q^{2})^{2}})(1+\frac{L_{1}^{2}}{r^{2}})]^{\frac{1}{2}}
[E_{2}^{2}-(1\\\label{12}&-&\frac{2mr^{2}}{(r^{2}+q^{2})^{\frac{3}{2}}}+
\frac{q^{2}r^{2}}{(r^{2}+q^{2})^{2}})(1+\frac{L_{2}^{2}}
{r^{2}})]^{\frac{1}{2}}-\frac{L_{1}L_{2}}{r^{2}}.
\end{eqnarray}
For $L=0$ in Eq.(\ref{8}), we have $V_{eff}(r)=Y(r)$ and hence at
$r=r_{min}=0.6q$, it gives $V_{eff}(r_{min})=Y(r_{min})$, which
leads to
\begin{equation}\label{13}
V_{eff}(r_{min})=1.19-0.453\frac{m}{q}=Y(r_{min}).
\end{equation}

We consider the collision of particles whose motion is along radial
geodesics with conserved energy $E_{1}=E_{2}=E=1$ (implies that the
ingoing particles asymptotically approach to the center) and angular
momentum is taken to be $L_{1}=L_{2}=L=0$ (particles reach the
center). Consequently, Eq.(\ref{12}) yields
\begin{equation}\label{14}
E^{2}_{cm}=\frac{4m^{2}}{(1-\frac{2mr^{2}}{(r^{2}+q^{2})^{\frac{3}{2}}}+
\frac{q^{2}r^{2}}{(r^{2}+q^{2})^{2}})}.
\end{equation}
The CME depends upon the chosen location of the colliding particles.
This will be maximum when the effective potential is minimum.
Inserting the value of $r_{min}$ in Eq.(\ref{14}), it follows that
\begin{equation}\label{15}
E^{2}_{cm,max}=\frac{4m^{2}}{1.19-0.453\frac{m}{q}}.
\end{equation}
This shows that the maximum CME depends on the ratio of mass and
charge of the regular BH, which is large for $q=0.4m$. When charge
is greater than mass by infinitesimally small amount, we introduce a
new parameter $\epsilon$ as
\begin{equation}\label{16}
\epsilon=1.19-.45\frac{m}{q}.
\end{equation}
For $\epsilon\rightarrow0$, the CME becomes infinite, i.e.,
\begin{equation}\label{17}
\lim_{\epsilon\rightarrow0}E^{2}_{cm,max}=\frac{4m^{2}}
{\epsilon}\rightarrow\infty.
\end{equation}

When we take a collision between particles with $E_{1}=E_{2}=E$ and
angular momentum $L_{1}=L_{2}=0$ moving along the same path, then
\begin{equation}\label{18}
\frac{E^{2}_{cm}}{2m^{2}}=\frac{2E^{2}}{(1-\frac{2mr^{2}}{(r^{2}+q^{2})^{\frac{3}{2}}}
+\frac{q^{2}r^{2}}{(r^{2}+q^{2})^{2}})},
\end{equation}
\begin{equation}\label{19}
U^{r}=\pm\sqrt{E^{2}-(1-\frac{2mr^{2}}{(r^{2}+q^{2})^{\frac{3}{2}}}
+\frac{q^{2}r^{2}}{(r^{2}+q^{2})^{2}})}.
\end{equation}
Since $Y(r_{min})=\epsilon=1.19-0.45\frac{m}{q}$, therefore, the
condition for $U^{r}$ to be real leads to
\begin{equation*}
E^{2}-Y(r_{min})\geq0 \quad\Rightarrow \quad E\geq\sqrt{\epsilon}.
\end{equation*}
If $E=\sqrt{\epsilon}$, then $U^{r}=0$, i.e., the particle remains
at rest for $r=r_{min}=0.6q$. If $E>\sqrt{\epsilon}$, then $U^{r}$
is real, i.e., an ingoing particle will come out as an outgoing
particle either after getting bounced back or after crossing over
the center. Thus the CME near the extremal limits is given by
\begin{equation}\label{20}
\lim_{\epsilon\rightarrow0}\frac{E^{2}_{cm}}{2m^{2}}=\frac{2E^{2}}{\epsilon}\rightarrow\infty.
\end{equation}

Finally, we consider the collision between an ingoing particle with
finite radial velocity as well as energy say $E_{1}$ and particle at
rest, where $r=r_{min}$ (having energy $E_{2}=\sqrt{\epsilon}$ and
zero angular momentum). The CME of these colliding particles is
\begin{equation}\label{21}
\frac{E^{2}_{min}}{2m^{2}}=1+\frac{E_{1}}{\sqrt{\epsilon}}.
\end{equation}
For extremal limits, this leads to
\begin{equation}\label{22}
\lim_{\epsilon\rightarrow0}\frac{E^{2}_{cm}}{2m^{2}}
=1+\frac{E_{1}}{\sqrt{\epsilon}}\approx\frac{E_{1}}
{\sqrt{\epsilon}}\rightarrow\infty,
\end{equation}
which is divergent. Hence the high energy collisions are independent
of horizon and singularities and can occur in the regular BH.

\section{Particle Acceleration in Charged Accelerating and Rotating Black Holes}

Here we investigate acceleration and particle collision in charged
accelerating and rotating BHs (non-extremal). Plebanski and
Demianski (PD) presented a class of type D BHs known as the family
of PD BHs (Plebanski and Demianski 1976). The general form of the
metric is
\begin{equation}\label{23}
ds^{2}=-f(r)dt^{2}+\frac{1}{g(r)}dr^{2}-2H(r)dtd\phi+\Sigma(r)d\theta^{2}+K(r)d\phi^{2}.
\end{equation}
We consider a PD BH (Sharif and Javed 2012) with zero NUT parameter
so that the metric is
\begin{equation}\label{24}
ds^{2}=-\frac{1}{\Omega^{2}}\{\frac{Q}{\rho^{2}}(dt-a\sin^{2}\theta
d\phi)^{2}-\frac{\rho^{2}}{Q}dr^{2}-\frac{\tilde{P}}{\rho^{2}}
(adt-(r^{2}+a^{2})d\phi)^{2}-\frac{\rho^{2}}{\tilde{P}}\sin^{2}\theta
d\phi^{2}\},
\end{equation}
with
\begin{eqnarray}\nonumber
\Omega&=&1-\frac{\alpha}{\omega}a\cos\theta r, \quad
\rho^{2}=r^{2}+a^{2}\cos^{2}\theta, \quad
\tilde{P}=P\sin^{2}\theta\\\nonumber
Q&=&(\omega^{2}k+e^{2}+g^{2}-2Mr+\frac{\omega^{2}k}{a^{2}}r^{2})(1-\frac{\alpha
a}{\omega}r)(1+\frac{\alpha a}{\omega}r).
\end{eqnarray}
Here $M$ and $a$ represent the mass and rotation of BH respectively,
while the parameters $e$ and $g$ are the electric and magnetic
charges, respectively, $\alpha$ represents acceleration of the BH.
The rotation parameter $\omega$ in terms of $a$ and $k$ is given by
$\frac{\omega^{2}}{a^{2}}k = 1$. It is interesting to mention here
that all the parameters $\alpha$, $M$, $e$, $g$ and $a$ vary
independently but $\omega$ depends on the value of $a$. For $\alpha
= 0$, the metric reduces to the Kerr-Newman BH, while $a = 0$ leads
to C-metric. In the limit $a=0=\alpha$, this yields RN BH, while for
$e=0=g$, the Schwarzschild BH can be obtained.

The horizons are found for $g(r)=0$, leading to
\begin{equation*}
\omega^{2}k+e^{2}+g^{2}-2Mr+\frac{\omega^{2}k}{a^{2}}r^{2}=0,
\end{equation*}
which is quadratic in $r$ with roots
\begin{equation*}
r_{\pm}= \frac{a^{2}}{\omega^{2}k}\{M\pm\sqrt{M^{2}-
\frac{\omega^{2}k}{a^{2}}(\omega^{2}k+e^{2}+g^{2})}\}.
\end{equation*}
For the existence of horizon
\begin{equation}\label{25}
M^{2}\geq \frac{\omega^{2}k}{a^{2}}(\omega^{2}k+e^{2}+g^{2}),
\end{equation}
where $r_{\pm}$ represent the inner and outer horizons. Also,
$r_{\alpha_{1}}= \frac{\omega}{\alpha a},~
r_{\alpha_{2}}=-\frac{\omega}{\alpha a}$ are acceleration horizons.
The angular velocity at outer horizon is
$\Omega_{H}=\frac{-g_{t\phi}}{g_{\phi \phi}},$ which provides
\begin{equation}\label{26}
\Omega_{H}=
\frac{a}{r_{+}^{2}+a^{2}}=\frac{a}{[\frac{a^{2}}{\omega^{2}k}
\{M+\sqrt{M^{2}-\frac{\omega^{2}k}{a^{2}}(\omega^{2}k+e^{2}+g^{2})}\}]^{2}+a^{2}}.
\end{equation}

The conserved energy and angular momentum along the geodesics are
\begin{eqnarray}\label{27}
E&=&(\frac{Q}{r^{2}}-\frac{Pa^{2}}{r^{2}})U^{t}
+(\frac{Pa(r^{2}+a^{2})}{r^{2}}-\frac{Qa}{r^{2}})U^{\phi},\\\label{28}
L&=&-(\frac{Pa(r^{2}+a^{2})}{r^{2}}-\frac{Qa}{r^{2}})U^{t}
+(\frac{P(r^{2}+a^{2})^{2}}{r^{2}}-\frac{Qa^{2}}{r^{2}})U^{\phi}.
\end{eqnarray}
These lead to the components of four velocity as follows
\begin{eqnarray}\label{29}
U^{t}&=&\frac{1}{PQ}\{E(\frac{P(r^{2}+a^{2})^{2}}{r^{2}}-\frac{Qa^{2}}{r^{2}})
-L(\frac{Pa(r^{2}+a^{2})}{r^{2}}-\frac{Qa}{r^{2}})\},\\\label{30}
U^{\phi}&=&\frac{1}{PQ}\{E(\frac{Pa(r^{2}+a^{2})}{r^{2}}-\frac{Qa}{r^{2}})
+L(\frac{Q}{r^{2}}-\frac{Pa^{2}}{r^{2}})\}.
\end{eqnarray}
The radial velocity component can be found using normalization
condition
\begin{eqnarray}\nonumber
U^{r}&=&\pm[\frac{1}{Pr^{4}}\{-PQ+E^{2}(P(r^{2}+a^{2})^{2}-Qa^{2})-L^{2}(Q-Pa^{2})\\\label{31}
&-&2EL(Pa(r^{2}+a^{2})-Qa)\}]^{\frac{1}{2}}.
\end{eqnarray}
Here $\pm$ correspond to radially ingoing and outgoing particles,
respectively. We introduce the effective potential as
\begin{equation}\label{32}
U^{r^{2}}+V_{eff}(r)=0,
\end{equation}
where
\begin{eqnarray}\nonumber
V_{eff}(r)&=&\frac{1}{Pr^{4}}\{PQ-E^{2}(P(r^{2}+a^{2})^{2}-Qa^{2})+L^{2}(Q-Pa^{2})\\\label{33}
&+&2EL(Pa(r^{2}+a^{2})-Qa)\}.
\end{eqnarray}
The conditions for circular orbit are
\begin{equation*}
V_{eff}(r)=0, \quad \frac{dV_{eff}(r)}{dr}=0.
\end{equation*}
We know that the timelike component of four velocity is greater than
zero $(U^{t}\geq0)$, yielding (using Eq.(\ref{29}))
\begin{equation}\label{34}
E(\frac{P(r^{2}+a^{2})^{2}}{r^{2}}-\frac{Qa^{2}}{r^{2}}) \geq
L(\frac{Pa(r^{2}+a^{2})}{r^{2}}-\frac{Qa}{r^{2}}).
\end{equation}
which reduces (at horizon) to $E\geq \frac{aL}{r^{2}_{+}+a^{2}}$.
The angular velocity of the BH (at $r=r_{+}$) is $
\Omega_{H}=\frac{a}{r^{2}_{+}+a^{2}}$. These lead to
\begin{equation}\label{35}
E\geq \Omega_{H}L.
\end{equation}

Now we explore the CME for two particles colliding with rest masses
$m_{1}$ and $m_{2}$ moving in equatorial motion. The CME of these
particles in the charged accelerating and rotating BH is given by
\begin{equation}\label{36}
\frac{E_{cm}}{\sqrt{2m_{1}m_{2}}}=\sqrt{\frac{(m_{1}
-m_{2})^{2}}{2m_{1}m_{2}}+\frac{M(r)-N(r)}{T(r)}},
\end{equation}
where
\begin{eqnarray*}
M(r)&=&\frac{1}{r^{6}}(E_{1}L_{2}+E_{2}L_{1})[Pa(r^{2}+a^{2})-Qa]
[2a^{2}P^{2}(r^{2}+a^{2})^{2}\\
&+&2a^{2}Q^{2}-PQ(r^{2}+2a^{2})^{2}]-[PQ+\frac{2}{r^{4}}(Pa(r^{2}+a^{2})-Qa)^{2}]\\
&\times&[\frac{E_{1}E_{2}}{r^{2}}(P(r^{2}+a^{2})^{2}-Qa^{2})-\frac{L_{1}L_{2}}{r^{2}}(Q-Pa^{2})],\\
N(r)&=&\sqrt{n_{1}(r)n_{2}(r)},\\
n_i(r)&=&E_{i}^{2}[\frac{P}{r^{2}}(r^{2}+a^{2})^{2}-\frac{Qa^{2}}{r^{2}}]
[\frac{2a^{2}}{r^{4}}(P(r^{2}+a^{2})+Q)^{2}\\
&+&PQ]+[L_{i}^{2}(\frac{Q}{r^{2}}-\frac{Pa^{2}}{r^{2}})+2E_{i}L_{i}
(\frac{P}{r^{2}}a(r^{2}+a^{2})-\frac{Qa}{r^{2}})]\\
&\times&[\frac{-2a^{2}}{r^{4}}(P(r^{2}+a^{2})+Q)^{2}+PQ],\\
T(r)&=&P^{2}[\omega^{2}k+e^{2}+g^{2}-2Mr+r^{2}]^{2}[1+\frac{\alpha
a}{\omega}r]^{2}[1-\frac{\alpha a}{\omega}r]^{2}.
\end{eqnarray*}
Here $E_i$ and $L_i$ are the conserved energy and angular momentum
for the $i$th particle. The above equations indicate that CME
depends on rotation.

For the CME near the horizon, i.e., for $r\rightarrow r_{+}$, the
term on the right side of Eq.(\ref{36}) is undetermined. Using
l'Hospital rule, it yields
\begin{equation}\label{37}
\frac{E_{cm}}{\sqrt{2m_{1}m_{2}}}=\sqrt{\frac{(m_{1}
-m_{2})^{2}}{2m_{1}m_{2}}+\frac{M'(r)-N'(r)}{T'(r)}},
\end{equation}
with
\begin{eqnarray*}
M'(r)\mid_{r=r_{+}}&=&-[PQ'-\frac{8Pa(r_{+}^{2}+a^{2})}{r_{+}^{5}}
+\frac{4Pa(r_{+}^{2}+a^{2})(2Par_{+}-Q')}{r_{+}^{4}}]\\
&\times&[\frac{E_{1}E_{2}P^{2}(r_{+}^{2}+a^{2})^{2}}{r_{+}^{2}}]+\frac{2P^{2}a^{2}(r_{+}^{2}
+a^{2})^{2}}{r_{+}^{4}}[\frac{E_{1}E_{2}}{r_{+}^{3}}(2P(r_{+}^{2}+a^{2})^{2}\\
&-&4Pr_{+}^{2}(r_{+}^{2}+a^{2})+Q'a^{2})+\frac{L_{1}L_{2}}{r_{+}^{3}}
(2Pa^{2}+r_{+}Q')]-[E_{1}L_{2}\\
&+&E_{2}L_{1}][\frac{6}{r_{+}^{7}}(2P^{3}a^{3}(r_{+}^{2}+a^{2})^{3})-(2a^{2}P^{2}
(r_{+}^{2}+a^{2})^{2})(2Par_{+}\\
&-&Q'a)+(Pa(r_{+}^{2}+a^{2}))(PQ'(r_{+}^{2}+2a^{2})^{2}-8a^{2}P^{2}r_{+}(r_{+}^{2}+a^{2}))],\\
N'(r)\mid_{r=r_{+}}&=&\frac{1}{2\sqrt{n_{1}(r_{+})n_{2}(r_{+})}}\{n'_{1}(r_{+})n_{2}(r_{+})+n'_{2}(r_{+})n_{1}(r_{+})\},
\end{eqnarray*}
\begin{eqnarray*}
n'_{i}(r)\mid_{r=r_{+}}&=&E_{i}^{2}\{[\frac{2a^{2}P^{2}
(r_{+}^{2}+a^{2})^{2}}{r_{+}^{4}}][\frac{4P(r_{+}^{2}+a^{2})}{r_{+}}
-\frac{2P(r_{+}^{2}+a^{2})^{2}}{r_{+}^{3}}-\frac{Q'a^{2}}{r_{+}^{2}}]\\
&+&[\frac{P(r_{+}^{2}+a^{2})^{2}}{r_{+}^{2}}][\frac{4a^{2}P(r_{+}^{2}+a^{2})
(2Pr_{+}+Q')}{r_{+}^{4}}-\frac{8a^{2}P^{2}(r_{+}^{2}+a^{2})^{2}}{r_{+}^{5}}\\
&+&PQ']\}-[\frac{2a^{2}P^{2}(r_{+}^{2}+a^{2})^{2}}{r_{+}^{4}}][L_{i}^{2}
(\frac{Q'}{r_{+}^{2}}+\frac{2Pa^{2}}{r_{+}^{3}})+2E_{i}L_{i}(\frac{2Pra}{r_{+}^{2}}\\
&-&\frac{Q'a}{r_{+}^{2}}-\frac{2Pa(r_{+}^{2}+a^{2})}{r_{+}^{3}})]
+[2E_{i}L_{i}\frac{Pa(r_{+}^{2}+a^{2})}{r_{+}^{2}}-L_{i}^{2}\frac{Pa^{2}}{r_{+}^{2}}]\\
&\times&[\frac{8a^{2}P(r_{+}^{2}+a^{2})^{2}}{r_{+}^{5}}+PQ'
-\frac{4a^{2}P(r_{+}^{2}+a^{2})}{r_{+}^{4}}(2Pr_{+}+Q')],\\
T'(r)\mid_{r=r_{+}}&=&2P^{2}QQ'\mid_{r=r_{+}}=0.
\end{eqnarray*}
For $m_{1}=m_{2}=m$, the value of $E^{2}_{cm}$ at $r_{+}$ turns out
to be
\begin{equation}\label{38}
\lim_{r\rightarrow r_{+}}\frac{E_{cm}^{2}}{2m^{2}}=\infty.
\end{equation}
Thus the CME of colliding particles for charged accelerating and
rotating BH moving in equatorial plane is infinite for the limiting
case.

For $a=0$, we have
\begin{eqnarray}\nonumber
E_{cm}&=&\sqrt{2}m\{1+\frac{E_{1}E_{2}r^{2}}{Q}-\frac{L_{1}L_{2}}{Pr^{2}}
-\frac{1}{Q}(-Q+E_{1}^{2}r^{2}-\frac{L_{1}^{2}Q}{Pr^{2}})^{\frac{1}{2}}(-Q\\\label{39}
&+&E_{2}^{2}r^{2}-\frac{L_{2}^{2}Q}{Pr^{2}})^{\frac{1}{2}}\}^{\frac{1}{2}}.
\end{eqnarray}
Expanding this at $Q=0$, it follows that
\begin{equation*}
E_{cm}=2m\sqrt{1+\frac{L_{1}-L_{2}}{4Pr^{2}}}
\end{equation*}
which is the $E_{cm}$ of the extremal Kerr-Newman BH with $a=0$.

\section{Conclusion}

In order to discuss the nature of matter and energy in particles
collision, the study of particle accelerators is of great
significance. It is believed that BH can behave as a particle
accelerator. In this paper, we have calculated the CME by using
equation of motion of particles moving in equatorial plane. We have
discussed particles colliding near the horizon of Ayon and Garcia BH
(regular BH) and Plebanski and Demianski BH with zero NUT (charged
accelerating and rotating BH). The CME of the two colliding
particles can be arbitrarily high, i.e., the collision can produce
energetic particles. For regular black hole, we have taken different
values of energy and angular momentum (conserved) and found that
unlike the common belief, high energy collision between the
particles can take place in a perfectly regular BH. It is found that
CME depends on the mass to charge ratio and can become unlimited for
some appropriate values of the parameters $(m,q)$. For the charged
rotating and accelerating BH, the CME turns out to be arbitrarily
high, which depends on rotation parameter only. We conclude that
center of mass energy turns out to be infinitely large for both BHs.
For $\alpha=0$, we have CME for the Kerr-Newmann BH and for
$a=0=\alpha$, results reduce to CME for the RN BH.

\end{document}